# Recent Developments in Nonregular Fractional Factorial Designs

Hongquan Xu, Frederick K. H. Phoa and Weng Kee Wong

*University of California, Los Angeles*

October 25, 2018

*Abstract:* Nonregular fractional factorial designs such as Plackett-Burman designs and other orthogonal arrays are widely used in various screening experiments for their run size economy and flexibility. The traditional analysis focuses on main effects only. Hamada and Wu (1992) went beyond the traditional approach and proposed an analysis strategy to demonstrate that some interactions could be entertained and estimated beyond a few significant main effects. Their groundbreaking work stimulated much of the recent developments in design criterion creation, construction and analysis of nonregular designs. This paper reviews important developments in optimality criteria and comparison, including projection properties, generalized resolution, various generalized minimum aberration criteria, optimality results, construction methods and analysis strategies for nonregular designs.

## 1 Introduction

In many scientific investigations, the main interest is in the study of effects of many factors simultaneously. Factorial designs, especially two-level or three-level factorial designs, are the most commonly used experimental plans for this type of investigation. A full factorial experiment allows all factorial effects to be estimated independently. However, it is often too costly to perform a full factorial experiment, so a fractional factorial design, which is a subset or fraction of a full factorial design, is preferred since it is cost-effective.

Fractional factorial designs are classified into two broad types: *regular* designs and *nonregular* designs. Regular designs are constructed through defining relations among factors and are described in many textbooks such as Box, Hunter and Hunter (2005), Dean and Voss (1999), Montgomery (2005) and Wu and Hamada (2000). These designs have a simple aliasing structure in that any two effects are either orthogonal or fully aliased. The run sizes are always a power of 2, 3 or a prime, and thus the "gaps" between possible run sizes are getting wider as the power increases. The concept of *resolution* (Box and Hunter 1961) and its refinement *minimum aberration* (Fries and Hunter 1980) play a pivotal role in the optimal choice of regular designs. There are many recent developments on minimum aberration designs; see Wu and Hamada (2000) and Mukerjee and Wu (2006) for further references.

Nonregular designs such as Plackett-Burman designs and other orthogonal arrays are widely used in various screening experiments for their run size economy and flexibility (Wu and Hamada, 2000). They fill the gaps between regular designs in terms of various run sizes and are flexible in accommodating various combinations of factors with different numbers of levels. Unlike regular designs, nonregular designs may exhibit a complex aliasing structure, that is, a large number of effects may be neither orthogonal nor fully aliased, which makes it difficult to interpret their significance. For this reason, nonregular designs were traditionally used to estimate factor main effects only but not their interactions. However, in many practical situations it is often questionable whether the interaction effects are negligible.

Hamada and Wu (1992) went beyond the traditional approach and proposed an analysis strategy to demonstrate that some interactions could be entertained and estimated through their complex aliasing structure. They pointed out that ignoring interactions can lead to (i) important effects being missed, (ii) spurious effects being detected, and (iii) estimated effects having reversed signs resulting in incorrectly recommended factor levels.

Much of the recent studies in nonregular designs were motivated from results in Hamada and Wu (1992). They included proposal of new optimality criteria, construction and analysis of nonregular designs. The primary aim of this paper is to review major developments in nonregular fractional factorial designs since 1992.

Here is a brief history of the major developments in nonregular designs. Plackett and Burman (1946) gave a large collection of two-level and three-level designs for multi-factorial experiments. These designs are often referred to as the Plackett-Burman designs in the literature. Rao (1947) introduced the concept of orthogonal arrays, including Plackett-Burman designs as special cases. Cheng (1980) showed that orthogonal arrays are universally optimal for main effects model. Hamada and Wu (1992) successfully demonstrated that some interactions could be identified beyond a few significant main effects for Plackett-Burman designs and other orthogonal arrays. Lin and Draper (1992) studied the geometrical projection properties of Plackett-Burman designs while Wang and Wu (1995) and Cheng (1995, 1998) studied the hidden projection properties of Plackett-Burman designs and other orthogonal arrays. The hidden projection properties provide an explanation for the success of the analysis strategy due to Hamada and Wu (1992). Sun and Wu (1993) were the first to coin the term "nonregular designs" when studying statistical properties of Hadamard matrices of order 16. Deng and Tang (1999) and Tang and Deng (1999) introduced the concepts of generalized resolution and generalized minimum aberration for two-level nonregular designs. Xu and Wu (2001) proposed the generalized minimum aberration for mixed-level nonregular designs. Because of the popularity of minimum aberration, the research on nonregular designs has been largely focused on the construction and properties of generalized minimum aberration designs. Our reference list suggests that keen interest in nonregular designs began in 1999 and continues to this day as evident by the increasing number of scientific papers on nonregular designs in major statistical journals.

Section 2 reviews the data analysis strategies for nonregular designs. Section 3 discusses the geometrical and hidden projection properties of the Plackett-Burman designs and other orthogonal arrays. Section 4 introduces the generalized resolution and generalized minimum aberration and their statistical justifications. Section 5 introduces the minimum moment aberration criterion, another popular criterion for nonregular designs. Section 6 considers uniformity and connections with various optimality criteria. Section 7 reviews construction methods and optimality results. Section 8 gives concluding remarks and future directions.

## 2 Analysis Strategies

We begin with a review of a breakthrough approach (Hamada and Wu 1992) by entertaining interactions in Plackett-Burman designs and other orthogonal arrays after identifying a few important main effects. Then we review another strategy proposed by Cheng and Wu (2001) for the dual purposes of factor screening and response surface exploration (or interaction detection) with quantitative factors.

The analysis strategy proposed by Hamada and Wu (1992) consists of three steps.

Step 1. Entertain all the main effects and interactions that are orthogonal to the main effects. Use standard analysis methods such as ANOVA and half-normal plots to select significant effects.

Step 2. Entertain the significant effects identified in the previous step and the two-factor interactions that consist of at least one significant effect. Identify significant effects using a forward selection regression procedure.

Step 3. Entertain the significant effects identified in the previous step and all the main effects. Identify significant effects using a forward selection regression procedure.

Iterate between Steps 2 and 3 until the selected model stops changing. Note that the traditional analysis of Plackett-Burman or other nonregular designs ends at Step 1.

Hamada and Wu (1992) based their analysis strategy on two empirical principles, *effect sparsity* and *effect heredity* (see Wu and Hamada 2000, Section 3.5). Effect sparsity implies that only few main effects and even fewer two-factor interactions are relatively important in a factorial experiment. Effect heredity means that in order for an interaction to be significant, at least one of its parent factors should be significant. Effect heredity excludes models that contain an interaction but none of its parent main effects, which lessens the problem of obtaining uninterpretable models. Hamada and Wu (1992) wrote that the strategy works well when both principles hold and the correlations between partially aliased effects are small to moderate. The effect sparsity suggests that only a few iterations will be required.

Using this procedure, Hamada and Wu (1992) reanalyzed data from three real experiments, a cast fatigue experiment using a 12-run Plackett-Burman design with seven 2-level factors, a blood glucose experiment using an 18-run mixed-level orthogonal array with one 2-level and seven 3-level factors, and a heat exchange experiment using a 12-run Plackett-Burman design with ten 2-level factors. They demonstrated that the traditional main effects analysis was limited and the results were misleading.

For illustration, consider the cast fatigue experiment conducted by Hunter, Hodi and Eager (1982) that used a 12-run Plackett-Burman design to study the effects of seven factors ($A$–$G$) on the fatigue life of weld repaired castings. Table 1 gives the data matrix and responses, where columns 8–11 are not used. The original analysis by Hunter, Hodi and Eager (1982) identified two significant factors $F$ and $D$. The factor $D$ had a much smaller effect with a $p$ value around 0.2. The fitted model was

$$\hat{y} = 5.73 + 0.458F - 0.258D, \qquad (1)$$

with a $R^2 = 0.59$. However, Hunter, Hodi and Eager (1982) noted a discrepancy between their fitted model (1) and previous work, namely, the sign of factor $D$ was reversed. Applying the three-step analysis strategy, Hamada and Wu (1992) identified a significant two-factor interaction $FG$ and obtained the following model

$$\hat{y} = 5.73 + 0.458F - 0.459FG. \qquad (2)$$

Table 1: Design Matrix and Responses, Cast Fatigue Experiment

| Run | Factor | | | | | | | | | | | Logged Lifetime |
| --- | --- | --- | --- | --- | --- | --- | --- | --- | --- | --- | --- | --- |
| | A | B | C | D | E | F | G | 8 | 9 | 10 | 11 | |
| 1 | + | + | − | + | + | + | − | − | − | + | − | 6.058 |
| 2 | + | − | + | + | + | − | − | − | + | − | + | 4.733 |
| 3 | − | + | + | + | − | − | − | + | − | + | + | 4.625 |
| 4 | + | + | + | − | − | − | + | − | + | + | − | 5.899 |
| 5 | + | + | − | − | − | + | − | + | + | − | + | 7.000 |
| 6 | + | − | − | − | + | − | + | + | − | + | + | 5.752 |
| 7 | − | − | − | + | − | + | + | − | + | + | + | 5.682 |
| 8 | − | − | + | − | + | + | − | + | + | + | − | 6.607 |
| 9 | − | + | − | + | + | − | + | + | + | − | − | 5.818 |
| 10 | + | − | + | + | − | + | + | + | − | − | − | 5.917 |
| 11 | − | + | + | − | + | + | + | − | − | − | + | 5.863 |
| 12 | − | − | − | − | − | − | − | − | − | − | − | 4.809 |

This model has $R^2 = 0.89$, which is a significant improvement over model (1) in terms of goodness of fit. The identification of $FG$ was not only consistent with the engineering knowledge reported in Hunter, Hodi and Eager (1982) but also provided a sound explanation on the discrepancy of the sign of factor $D$. The coefficient of $D$ in (1) actually estimates $D + \frac{1}{3}FG$ and therefore the sign of $D$ in (1) could be negative even if $D$ had a small positive effect. This experiment was later reanalyzed with other methods by several authors, including Box and Meyer (1993), Chipman, Hamada and Wu (1997), Westfall, Young and Lin (1998), Yuan, Joseph and Lin (2007), and Phoa, Pan and Xu (2007).

Hadama and Wu (1992) discussed limitations of their analysis strategy and provided solutions. Wu and Hamada (2000, chap. 8) further suggested some extensions such as the use of all subset variable selection if possible.

For quantitative factors with more than two levels, Cheng and Wu (2001) proposed the following two-stage analysis strategy to achieve the dual objectives of factor screening and response surface exploration (or interaction detection) using a single design. This two-stage analysis strategy is also the two key aspects in standard response surface methodology.

Stage 1. Perform factor screening and identify important factors.

Stage 2. Fit a second-order model for the factors identified in stage 1.

For $m$ quantitative factors, denoted by $x_1, \ldots, x_m$, the second-order model is

$$y = \beta_0 + \sum_{i=1}^{m} \beta_i x_i + \sum_{i=1}^{m} \beta_{ii} x_i^2 + \sum_{1=i<j}^{m} \beta_{ij} x_i x_j + \epsilon,$$

where $\beta_0, \beta_i, \beta_{ii}, \beta_{ij}$ are unknown parameters and $\epsilon$ is the error term.

Cheng and Wu (2001) proposed that the two-stage analysis be broken down into three parts: screening analysis in stage 1, projection that links stages 1 and 2, and response surface exploration in stage 2. Various screening analyses can be utilized in stage 1, such as the conventional ANOVA or half-normal plots on the main effects. Their analysis strategy again assumes that effect sparsity and effect heredity principles hold. They reanalyzed a PVC insulation experiment reported by Taguchi (1987) that used a regular 27-run design with nine 3-level factors. They identified a significant linear-by-linear interaction effect which was missed by Taguchi.

For illustration, consider an experiment reported by King and Allen (1987) that used an 18-run orthogonal array to study the effects of one two-level factor ($A$) and seven three-level factors ($B$–$H$) on radio frequency chokes. Each run had two replicates and Table 2 gives the design matrix and the responses. Xu, Cheng and Wu (2004) performed data analysis following the two-stage analysis strategy. At the first stage, they fitted an ANOVA model for main effects and found that four factors $B$, $E$, $G$, and $H$ were significant at the usual 5% level. At the second stage, they fitted a second-order model among the four active factors and obtained the following nine-effect model:

$$\hat{y} = 105.1 + 2.61B - 4.05E - 7.75G + 2.91H - 2.85E^2 + 1.39BE - 3.30EG - 1.41EH + 1.86GH,$$

where the levels 0, 1, 2 were coded as $-1, 0, 1$, respectively. The model has $R^2 = 0.96$, indicating a good fit. It is worthwhile to point out that the 18-run design does not have enough degrees of freedom to estimate all six two-factor interactions among four factors (since each two-factor interaction has four degrees of freedom).

More sophisticated analysis strategies have been proposed for experiments with complex aliasing. Box and Meyer (1993) proposed a Bayesian method for finding the active factors in screening experiments. Chipman, Hamada and Wu (1997) proposed a Bayesian approach that employs a Gibbs sampler to perform an efficient stochastic search of the model space. Many other recent variable selection methods can also be used for analyzing nonregular designs. For example, Yuan, Joseph and Lin (2007) suggested an extension of the general-purpose LARS (least angle regression), first proposed by Efron, Hastie, Johnstone and Tibshirani (2004).

Table 2: Design Matrix and Responses, Radio Frequency Chokes Experiment

| Run | A | B | C | D | E | F | G | H | Responses | |
|---:|---|---|---|---|---|---|---|---|---:|---:|
| 1 | 0 | 0 | 0 | 0 | 0 | 0 | 0 | 0 | 106.20 | 107.70 |
| 2 | 0 | 0 | 1 | 1 | 1 | 1 | 1 | 1 | 104.20 | 102.35 |
| 3 | 0 | 0 | 2 | 2 | 2 | 2 | 2 | 2 | 85.90 | 85.90 |
| 4 | 0 | 1 | 0 | 0 | 1 | 1 | 2 | 2 | 101.15 | 104.96 |
| 5 | 0 | 1 | 1 | 1 | 2 | 2 | 0 | 0 | 109.92 | 110.47 |
| 6 | 0 | 1 | 2 | 2 | 0 | 0 | 1 | 1 | 108.91 | 108.91 |
| 7 | 0 | 2 | 0 | 1 | 0 | 2 | 1 | 2 | 109.76 | 112.66 |
| 8 | 0 | 2 | 1 | 2 | 1 | 0 | 2 | 0 | 97.20 | 94.51 |
| 9 | 0 | 2 | 2 | 0 | 2 | 1 | 0 | 1 | 112.77 | 113.03 |
| 10 | 1 | 0 | 0 | 2 | 2 | 1 | 1 | 0 | 93.15 | 92.83 |
| 11 | 1 | 0 | 1 | 0 | 0 | 2 | 2 | 1 | 97.25 | 100.6 |
| 12 | 1 | 0 | 2 | 1 | 1 | 0 | 0 | 2 | 109.51 | 113.28 |
| 13 | 1 | 1 | 0 | 1 | 2 | 0 | 2 | 1 | 85.63 | 86.91 |
| 14 | 1 | 1 | 1 | 2 | 0 | 1 | 0 | 2 | 113.17 | 113.45 |
| 15 | 1 | 1 | 2 | 0 | 1 | 2 | 1 | 0 | 104.85 | 98.87 |
| 16 | 1 | 2 | 0 | 2 | 1 | 2 | 0 | 1 | 113.14 | 113.78 |
| 17 | 1 | 2 | 1 | 0 | 2 | 0 | 1 | 2 | 103.19 | 106.46 |
| 18 | 1 | 2 | 2 | 1 | 0 | 1 | 2 | 0 | 95.70 | 97.93 |

# 3 Projection Properties of Plackett-Burman Designs and Other Orthogonal Arrays

We first review the concept of orthogonal arrays due to Rao (1947). An orthogonal array of $N$ runs, $m$ factors, $s$ levels and strength $t$, denoted by $OA(N, s^m, t)$, is an $N \times m$ matrix in which each column has $s$ symbols or levels and for any $t$ columns all possible $s^t$ combinations of symbols appear equally often in the matrix. Rao (1973) generalized the definition to the asymmetrical case where an orthogonal array is allowed to have variable numbers of symbols, i.e., mixed levels. For example, the 12-run Plackett-Burman design in Table 1 is an $OA(12, 2^{11}, 2)$ and the 18-run design in Table 2 is an $OA(18, 2^1 3^7, 2)$. Hedayat, Sloane and Stufken (1999) gave a comprehensive description on various aspects of orthogonal arrays.

Plackett-Burman designs are saturated orthogonal arrays of strength two because all degrees of freedom are utilized to estimate main effects. Orthogonal arrays of strength two allow all the main effects to be estimated independently and they are universally optimal for the main effects model (Cheng 1980). A necessary condition for the existence of an $OA(N, s^m, 2)$ is that $N - 1 \geq m(s-1)$. A design is called saturated if $N - 1 = m(s - 1)$ and supersaturated if $N - 1 < m(s - 1)$. In the literature, orthogonal arrays of strength two are often called orthogonal designs or orthogonal arrays without mentioning the strength explicitly.

Orthogonal arrays include both regular and nonregular designs. For regular designs, the concepts of strength and resolution are equivalent because a regular design of resolution $R$ is an orthogonal array of strength $t = R - 1$. For a regular design of resolution $R$, the projection onto any $R$ factors must be either a full factorial or copies of a half-replicate of a full factorial. The projection for nonregular designs is more complicated.

Plackett-Burman designs are of strength two so that the projection onto any two factors is a full factorial. Lin and Draper (1992) studied the geometrical projection properties of the Plackett-Burman designs onto three or more factors. Their computer searches found all the projections of 12-, 16-, 20-, 24-, 28-, 32- and 36-run Plackett-Burman designs onto three factors. They found that these projections must have at least a copy of the full $2^3$ factorial or at least a copy of a $2^{3-1}$ replicate or both. In particular, any projection onto three factors must contain a copy of a full factorial except for the 16- and 32-run Plackett-Burman designs, which are regular designs. The important statistical implication of this finding is that if only at most three factors are truly important, then after identifying the active factors, all factorial effects among these active factors are estimable, regardless which three factors are important.

Box and Tyssedal (1996) defined a design to be of projectivity $p$ if the projection onto every subset of $p$ factors contains a full factorial design, possibly with some points replicated. It follows from these definitions that an orthogonal array of strength $t$ is of projectivity $t$. Cheng (1995) showed that, as long as the run size $N$ is not a multiple of $2^{t+1}$, an $OA(N, 2^m, t)$ with $m \geq t+2$ has projectivity $t+1$, even though the strength is only $t$.

The 12-run Plackett-Burman design given in Table 1 is of projectivity three but not of projectivity four. Wang and Wu (1995) found that its projection onto any four factors has the property that all the main effects and two-factor interactions can be estimated if the higher-order interactions are negligible. They referred this estimability of interactions without relying on geometric projection to as having a *hidden projection* property.

More generally, Wang and Wu (1995) defined a design as having a hidden projection property if it allows some or all interactions to be estimated even when the projected design does not have the right resolution or other geometrical/combinatorial design property for the same interactions to be estimated. For the Plackett-Burman designs their hidden projection property is a result of complex aliasing between the interactions and the main effects. For example, in the 12-run Plackett-Burman design given in Table 1, any two-factor interaction, say $AB$, is orthogonal to the main effects $A$ and $B$, and partially aliased with all other main effects with correlation $1/3$ or $-1/3$. Because no two-factor interaction is fully aliased with any main effects, it is possible to estimate four main effects and all six two-factor interactions among them together.

The general results on hidden projection properties were obtained by Cheng (1995, 1998) and Bulutoglu and Cheng (2003). Cheng (1995) showed that as long as the run size $N$ of an $OA(N, 2^m, 2)$ is not a multiple of 8, its projection onto any four factors allows the estimation of all the main effects and two-factor interactions when the higher-order interactions are negligible. Bulutoglu and Cheng (2003) showed that the same hidden projection property also holds for Paley designs [constructed by a method due to Paley (1933)] of sizes greater than 8, even when their run sizes are multiples of 8. A key result is that such designs do not have defining words of length three or four. Cheng (1998) further showed that as long as the run size $N$ of an $OA(N, 2^m, 3)$ is not a multiple of 16, its projection onto any five factors allows the estimation of all the main effects and two-factor interactions. Cheng (2006) gave a nice review of projection properties of factorial designs and their role in factor screening.

A few papers studied projection properties of designs with more than two levels. Wang and Wu (1995) studied the hidden projections onto 3 and 4 factors of the popular $OA(18, 3^7, 2)$ given in Table 2 (columns $B$–$H$). Cheng and Wu (2001) further studied the projection properties of this $OA(18, 3^7, 2)$ and an $OA(36, 3^{12}, 2)$ in terms of their two-stage analysis strategy. They constructed a nonregular $OA(27, 3^8, 2)$ that allows the second-order model to be estimated in all four-factor projections. In contrast, any regular 27-run design with eight 3-level factors does not have this four-factor projection property. They concluded that three-level nonregular designs have better projection properties and are more useful than regular designs for the dual purposes of factor screening and response surface exploration. Xu, Cheng and Wu (2004) further explored the projection properties of 18-run and 27-run orthogonal arrays and constructed a nonregular $OA(27, 3^{13}, 2)$ that allows the second-order model to be estimated in all of the five-factor projections. Tsai, Gilmour and Mead (2000, 2004), Evangelaras, Koukouvinos, Dean and Dingus (2005) and Evangelaras, Koukouvinos and Lappas (2007, 2008) also studied projection properties of three-level orthogonal arrays. Dey (2005) studied projectivity properties of asymmetrical orthogonal arrays with all except one factors having two levels.

## 4  Generalized Resolution and Generalized Minimum Aberration

Prior to 1999, an outstanding problem was how to assess, compare and rank nonregular designs in a systematic fashion. Deng and Tang (1999) and Tang and Deng (1999) were the first to propose generalized resolution and generalized minimum aberration criteria for 2-level nonregular designs, which are natural generalizations of the traditional concepts of resolution and minimum aberration for regular designs.

To define these two important concepts, generalized resolution and generalized minimum aberration, Deng and Tang (1999) and Tang and Deng (1999) introduced the important notion of $J$-characteristics and is defined as follows. Given a two-level $N \times m$ design $D = (d_{ij})$, for $s = \{c_1, \ldots, c_k\}$, a subset of $k$ columns of $D$, define

$$j_k(s) = \sum_{i=1}^{N} c_{i1} \cdots c_{ik} \text{ and } J_k(s) = |j_k(s)|, \qquad (3)$$

where $c_{ij}$ is the $i$th component of column $c_j$. The quantity $j_k(s)/N$ can be viewed as an extension of correlation. For illustration, consider the 12-run Plackett-Burman design given in Table 1. For $s = \{A, B\}$, $j_2(s) = 0$ since $A$ and $B$ are orthogonal. For $s = \{A, B, C\}$, $j_3(s)/N = -1/3$ is the correlation between main effect $A$ and two-factor interaction $BC$. For $s = \{A, B, C, D\}$, $j_4(s)/N = -1/3$ is the correlation between two two-factor interactions, say $AB$ and $CD$. The quantity $\rho_k(s) = J_k(s)/N$ is called the normalized $J$-characteristics by Tang and Deng (1999) or aliasing index by Cheng, Li and Ye (2004) and Phoa and Xu (2008) because $0 \leq \rho_k(s) \leq 1$. It is not difficult to see that if $D$ is a two-level regular design then $\rho_k(s) = 0$ or 1 for all $s$. Ye (2004) showed that the reserve is also true. Therefore, for a nonregular design, there always exist some $s$ such that $0 < \rho_k(s) < 1$.

Suppose that $r$ is the smallest integer such that $\max_{|s|=r} J_r(s) > 0$, where the maximization is over all subsets of $r$ columns. Then the *generalized resolution* is defined to be

$$R = r + \delta, \text{ where } \delta = 1 - \max_{|s|=r} \frac{J_r(s)}{N}. \tag{4}$$

For the 12-run design in Table 1, $r = 3$, $\delta = 2/3$ and the generalized resolution is $R = 3.67$. It is easy to see that for an $OA(N, 2^m, t)$, $j_k(s) = 0$ for any $k \leq t$ and therefore $r \leq R < r + 1$ where $r = t + 1$. If $\delta > 0$, a subset $s$ of $D$ with $r$ columns contains at least $N\delta/2^r$ copies of a full $2^r$ factorial and therefore the projectivity of $D$ is at least $r$ (Deng and Tang (1999)). For a regular design, $\delta = 0$ and the projectivity is exactly $r - 1$.

Two regular designs of the same resolution can be distinguished using the minimum aberration criterion, and the same idea can be applied to nonregular designs using the *minimum G-aberration* criterion (Deng and Tang (1999)). Roughly speaking, the minimum $G$-aberration criterion always chooses a design with the smallest confounding frequency among designs with maximum generalized resolution. Formally, the minimum $G$-aberration criterion is to sequentially minimize the components in the confounding frequency vector

$$\text{CFV}(D) = [(f_{11}, \ldots, f_{1N}); (f_{21}, \ldots, f_{2N}); \ldots; (f_{m1}, \ldots, f_{mN})],$$

where $f_{kj}$ denotes the frequency of $k$-column combinations $s$ with $J_k(s) = N + 1 - j$.

Minimum $G$-aberration is very stringent and it attempts to control $J$-characteristics in a very strict manner. Tang and Deng (1999) proposed a relaxed version of minimum $G$-aberration and called it the *minimum $G_2$-aberration* criterion. Let

$$A_k(D) = N^{-2} \sum_{|s|=k} J_k^2(s). \tag{5}$$

The vector $(A_1(D), \ldots, A_m(D))$ is called the *generalized wordlength pattern*, because for a regular design $D$, $A_k(D)$ is the number of words of length $k$ in the defining contrast subgroup of $D$. The *minimum $G_2$-aberration* criterion (Tang and Deng (1999)) is to sequentially minimize the generalized wordlength pattern $A_1(D), A_2(D), \ldots, A_m(D)$.

For regular designs both minimum $G$-aberration and minimum $G_2$-aberration criteria reduce to the traditional minimum aberration criterion. However, these two criteria can result in selecting different nonregular designs. We note that minimum $G$-aberration nonregular designs always have maximum generalized resolution whereas minimum $G_2$-aberration nonregular designs may not. This is in contrast to regular case where minimum aberration regular designs always have maximum resolution among all regular designs.

Tang and Deng (1999) also defined minimum $G_e$-aberration for any $e > 0$ by replacing $J_k^2(s)$ with $J_k^e(s)$ in (5). However, only the minimum $G_2$-aberration criterion is popular due to various statistical justifications and theoretical results.

Xu and Wu (2001) proposed the *generalized minimum aberration* criterion for comparing asymmetrical (or mixed-level) designs. The generalized minimum aberration criterion was motivated from ANOVA models and includes the minimum $G_2$-aberration criterion as a special case. By exploring an important connection between design theory and coding theory, Xu and Wu (2001) showed that the generalized wordlength pattern defined in (5) are linear combinations of the distribution of pairwise distance between the rows. This observation plays a pivotal role in the subsequent theoretical development of nonregular designs.

Ma and Fang (2001) independently extended the minimum $G_2$-aberration criterion for designs with more than two levels. They named their criterion as the *minimum generalized aberration* criterion, which is a special case of the generalized minimum aberration criterion proposed by Xu and Wu (2001).

Ye (2003) redefined the generalized wordlength pattern and generalized minimum aberration for two-level designs using indicator functions. Cheng and Ye (2004) defined generalized resolution and generalized minimum aberration criterion for quantitative factors. The generalized minimum aberration criterion proposed by Xu and Wu (2001) is independent of the choice of treatment contrasts and thus model-free whereas the generalized minimum aberration criterion by Cheng and Ye (2004) depends on the specific model.

### 4.1 Statistical Justifications

Deng and Tang (1999) provided a statistical justification for the generalized resolution by showing that designs with maximum generalized resolution minimize the contamination of nonnegligible two-factor interactions on the estimation of main effects. Tang and Deng (1999) provided a similar statistical justification for minimum $G_2$-aberration designs. In a further extension, Xu and Wu (2001) gave a statistical justification for generalized minimum aberration designs with mixed levels.

A common situation that arises in practice is that the main effects are of primary interest but there are uninteresting yet non-negligible interactions that we know will affect the main effects estimates. To fix ideas, consider a two-level $N \times m$ design $D = (d_{ij})$ with columns denoted by $d_1, \ldots, d_m$ and generalized resolution between 3 and 4. Suppose that one fits a main effects model

$$y_i = \beta_0 + \sum_{j=1}^{m} \beta_j d_{ij} + \epsilon_i, \tag{6}$$

but the true model is

$$y_i = \beta_0 + \sum_{j=1}^{m} \beta_j d_{ij} + \sum_{k<l}^{m} \beta_{kl} d_{ik} d_{il} + \epsilon_i. \tag{7}$$

The least squares estimator $\hat{\beta}_j$ of $\beta_j$ from the working model (6), under the true model (7), has expectation given by

$$E(\hat{\beta}_j) = \beta_j + N^{-1} \sum_{k<l}^{m} j_3(d_j, d_k, d_l) \beta_{kl}$$

for $j = 1, \ldots, m$, where $j_3(d_j, d_k, d_l)$ is defined in (3). There are many ways to minimize the biases in estimating main effects due to the presence of the interaction effects. A conservative approach is minimizing the maximum bias, $\max_{j<k<l} J_3(d_j, d_k, d_l)$. This is equivalent to maximizing the generalized resolution as defined in (4). Therefore, designs with maximum generalized resolution minimize the maximum bias of nonnegligible interactions on the estimation of the main effects. A more aggressive approach is minimizing the sum of squared coefficients $\sum_{j=1}^{m} \sum_{k<l}^{m} [j_3(d_j, d_k, d_l)/N]^2 = 3A_3(D)$, where $A_3(D)$ is defined in (5). Hence minimum $G_2$-aberration designs minimize the overall contamination of nonnegligible interactions on the estimation of the main effects.

For regular designs, Cheng, Steinberg and Sun (1999) justified the minimum aberration criterion by showing that it is a good surrogate for some model-robustness criteria. Following their approach, Cheng, Deng and Tang (2002) considered the situation where (i) the main effects are of primary interest and their estimates are required and (ii) the experimenter would like to have as much information about two-factor interactions as possible, under the assumption that higher-order interactions are negligible. Without knowing which two-factor interactions are significant, they considered the set of models containing all of the main effects and $f$ two-factor interactions for $f = 1, 2, 3, \ldots$. Let $E_f$ be the number of estimable models and $D_f$ be the average of D-efficiencies. Cheng, Deng and Tang (2002) showed that the minimum $G_2$-aberration designs tend to have large $E_f$ and $D_f$ values, especially for small $f$; therefore, the minimum $G_2$-aberration criterion provides a good surrogate for the traditional model-dependent efficiency criteria. Ai, Li and Zhang (2005) and Mandal and Mukerjee (2005) extended their approach to mixed-level designs.

## 5 Minimum Moment Aberration

Based on coding theory, Xu (2003) proposed the minimum moment aberration criterion for assessing nonregular designs. For an $N \times m$ design $D$ with $s$ levels and a positive integer $t$, define the $t$th power moment to be

$$K_t(D) = [N(N-1)/2]^{-1} \sum_{1 \leq i < j \leq N} [\delta_{ij}(D)]^t, \qquad (8)$$

where $\delta_{ij}(D)$ is the number of coincidences between the $i$th and $j$th rows. For two row vectors $(x_1, \ldots, x_m)$ and $(y_1, \ldots, y_m)$, the number of coincidences is the number of $i$'s such that $x_i = y_i$. Note that $m - \delta_{ij}(D)$ is known as the *Hamming distance* between the $i$th and $j$th rows in coding theory.

The power moments measure the similarity among runs (i.e., rows). The first and second power moments measure the average and variance of the similarity among runs. Minimizing the power moments makes runs to be as dissimilar as possible. Therefore, good designs should have small power moments. This leads to the *minimum moment aberration* criterion (Xu (2003)) that is to sequentially minimize the power moments $K_1(D), K_2(D), \ldots, K_m(D)$.

We note that the computation of the power moments involves the number of coincidences between rows. By applying generalized MacWilliams identities and Pless power moment identities, two fundamental results in coding theory (see, e.g., MacWilliams and Sloane 1977, chap. 5), Xu (2003) showed that the power moments $K_t$ defined in (8) are linear combinations of the wordlength patterns $A_1, \ldots, A_t$ in (5). Specifically,

$$K_t(D) = c_t A_t(D) + c_{t-1} A_{t-1}(D) + \ldots + c_1 A_1(D) + c_0,$$

where $c_i$ are constants depending on $i, N, m, s$ only and the leading coefficient $c_t$ is positive. It is not difficult to see now that sequentially minimizing $K_1(D), \ldots, K_m(D)$ is equivalent to sequentially minimizing $A_1(D), \ldots, A_m(D)$. Therefore, the minimum moment aberration is equivalent to the generalized minimum aberration.

The equivalence of the minimum moment aberration and the generalized minimum aberration is very important. On the one hand, it not only provides a geometrical justification for the generalized minimum aberration, but also provides a statistical justification for the minimum moment aberration. On the other hand, it provides a useful tool for efficient computation and theoretical development. For an $N \times m$ design with two levels, the complexity of computing the generalized wordlength pattern according to the definition (5) is $O(N2^m)$ whereas the complexity of computing $m$ power moments is $O(N^2 m^2)$. The saving in computation is tremendous when the number of factors $m$ is large. This observation led to successful algorithmic constructions of mixed-level orthogonal arrays (Xu (2002)), a catalog of 3-level regular designs (Xu (2005b)), and blocked regular designs with minimum aberration (Xu and Lau (2006)). As a theoretical tool, Xu (2003) developed a unified theory for nonregular and supersaturated designs. Xu and Lau (2006) and Xu (2006) further used the concept of minimum moment aberration to develop a theory for blocked regular designs and constructed minimum aberration blocked regular designs with 32, 64 and 81 runs.

To mimic the minimum $G$-aberration criterion (Deng and Tang (1999)), Xu and Deng (2005) applied the minimum moment aberration criterion to projection designs and proposed the *moment aberration projection* to rank and classify general nonregular designs. It was a surprise that the minimum $G$-aberration criterion and the moment aberration projection criterion are not equivalent for two-level designs. Xu and Deng (2005) provided examples to show that the latter is more powerful in classifying and ranking nonregular designs than the former. They also provided examples to illustrate that the moment aberration projection criterion is supported by other design criteria. The concept of moment projection turns out to be very useful in the algorithmic construction of regular designs; see Xu (2005b, 2007).

For mixed-level designs, Xu (2003) suggested to weight each column according to its level, called natural weights, and replace $\delta_{ij}(D)$ in (8) with the number of weighted coincidences. Xu (2003) showed that the minimum moment aberration with natural weights is weakly equivalent to the generalized minimum aberration for mixed-level designs.

## 6  Uniformity and Connection Among Various Criteria

Uniformity or space filling is a desirable design property for computer experiments (Fang, Li and Sudjianto (2006)). Various uniformity measures are used to assess the space filling property for the so-called uniform design (Fang and Wang (1994), Fang, Lin, Winker and Zhang (2000)). Fang and Mukerjee (2000) found a connection between aberration and uniformity for 2-level regular designs. This connection was extended by Ma and Fang (2001) for general two-level designs. The basic result states that for a two-level $N \times m$ design $D$, regular or nonregular, the centered $L_2$-discrepancy ($CL_2$), a uniformity measure introduced by Hickernell (1998), can be expressed in terms of its generalized wordlength pattern $A_k(D)$ as follows:

$$\{CL_2(D)\}^2 = \left(\frac{13}{12}\right)^m - 2\left(\frac{35}{32}\right)^m + \left(\frac{9}{8}\right)^m \left\{1 + \sum_{k=1}^{m} \frac{A_k(D)}{9^k}\right\}.$$

Since the coefficient of $A_k(D)$ decreases exponentially with $k$, one can anticipate that designs with small $A_k(D)$ for small values of $k$ should have small $\{CL_2(D)\}^2$; in other words, minimum $G_2$-aberration designs tend to be uniform over the design region. Ma and Fang (2001) also gave analytic formulas that link the generalized wordlength pattern with other uniformity measures for two- and three-level designs.

Tang (2001) showed that minimum $G_2$-aberration designs have good low-dimensional projection properties. Ai and Zhang (2004a) extended this result to mixed-level designs and showed that generalized minimum aberration designs have good low-dimensional projection properties.

There is much more work on the connection among aberration, uniformity and projection. Hickernell and Liu (2002) showed that generalized minimum aberration designs and minimum discrepancy designs are equivalent in a certain limit. Qin and Fang (2004), Ai, Li and Zhang (2005), Fang and Qin (2005), Liu, Fang and Hickernell (2006), Qin and Ai (2007), and Qin, Zou and Chatterjee (2008) discussed the connections among different criteria for symmetrical and asymmetrical fractional factorial designs, including generalized minimum aberration, minimum moment aberration, and various uniformity measures.

# 7 Construction and Optimality Results

An important and challenging issue is the construction of good nonregular designs. There are two simple reasons: (i) nonregular designs do not have a unified mathematical description; (ii) there are much more nonregular designs than regular designs. Since 1999, a main stream of researches focused on searching or constructing nonregular designs with good properties in terms of the minimum $G_2$-aberration and generalized minimum aberration criteria. This section reviews algorithmic constructions and optimality results. The last subsection reviews a simple yet powerful construction method via quaternary codes.

## 7.1 Algorithmic constructions

Two-level nonregular designs are often constructed from Hadamard matrices. A Hadamard matrix of order $N$ is an $N \times N$ matrix with the elements $\pm 1$ whose columns (and rows) are orthogonal to each other. From a Hadamard matrix of order $N$, one obtains a saturated two-level orthogonal array with $N$ runs and $N-1$ columns, which is a nonregular design if $N$ is not a power of 2. Neil Sloane of AT&T Shannon Labs maintains a large collection of Hadamard matrices at his website http://www.research.att.com/~njas/, which includes all Hadamard matrices of orders $N$ up through 28, and at least one of every order $N$ up through 256. Sloane also maintains a library of orthogonal arrays as a companion to the book by Hedayat, Sloane and Stufken (1999). SAS maintains a library of orthogonal arrays (of strength two) up through 144 runs at http://support.sas.com/techsup/technote/ts723.html. SAS also provides a set of free macros for constructing over 117,000 orthogonal arrays up through 513 runs, which are documented in the free Web book by Kuhfeld (2005).

A simple strategy for constructing generalized minimum aberration designs is searching over all possible projection designs from existing Hadamard matrices or orthogonal arrays. Deng and Tang (2002) presented a catalog of generalized minimum aberration designs by searching over Hadamard matrices of order 16, 20, and 24. However, limiting to Hadamard matrices may miss the optimal design in some cases; therefore, Li, Tang and Deng (2004) searched generalized minimum aberration designs from and outside Hadamard matrices with 20, 24, 28, 32 and 36 runs. They found that the best $20 \times 6$ and $20 \times 7$ designs according to minimum $G$-aberration cannot be obtained from Hadamard matrices. Similarly, Xu and Deng (2005) considered the construction of optimal designs under the moment aberration projection criterion. Besides searching over all Hadamard matrices of order 16 and 20, they searched over all projection designs from 68 saturated $OA(27, 3^{13}, 2)$ from Lam and Tonchev (1996). They also observed that not all 20-run and 27-run moment aberration projection designs can be embedded into Hadamard matrices or saturated orthogonal arrays.

Sun, Li and Ye (2002) proposed an algorithm for sequentially constructing non-isomorphic orthogonal designs. Two designs are said to be *isomorphic* or *equivalent* if one design can be obtained from the other by row permutations, column permutations, or relabeling of levels. An essential element of their algorithm is using minimal column base to reduce the computations for determining isomorphism between any two designs. By using this algorithm, they obtained the complete catalogs of two-level orthogonal designs for 12, 16, and 20 runs. Their results suggest that there is only one unique $12 \times m$ design for $m = 4$ and $7 \leq m \leq 11$ and that there are two non-isomorphic $12 \times m$ design for $m = 5$ and 6. All these designs can be found as projection designs of the 12-run Plackett-Burman design given in Table 1. They found that there are five $16 \times 15$ orthogonal designs, which are equivalent to the five non-isomorphic Hadamard matrices of order 16 by Hall (1961). An important result is that all 16-run orthogonal designs are projections of one of the five 16-run Hadamard matrices. They found that there are three $20 \times 19$ orthogonal designs, which are equivalent to the three non-isomorphic Hadamard matrices of order 20 by Hall (1965). From their complete catalog, they obtained generalized minimum aberration designs. They found that most of the generalized minimum aberration designs are projections of the 20-run Hadamard matrices and thus agrees with the designs reported in Deng and Tang (2002). However, they found the generalized minimum aberration designs for $m = 6$ and $m = 7$ are not projections of the Hadamard matrices. This agrees with the results from Li, Tang and Deng (2004) and Xu and Deng (2005). The complete catalogs of 12, 16 and 20 runs were later used by Li, Lin and Ye (2003) in the choice of optimal foldover plans, by Cheng, Li and Ye (2004) in the construction of blocked nonregular designs, by Loeppky, Bingham and Sitter (2006) for constructing nonregular robust parameter designs, and by Li (2006) for constructing screening designs for model selection.

Xu, Cheng and Wu (2004) considered the design issues related to the dual objectives of factor screening and interaction detection for quantitative factors. They proposed a set of optimality criteria to assess the performance of designs and a three-step approach to searching for optimal designs. They not only searched over all projection designs from the commonly used $OA(18, 3^7, 2)$ given by columns $B$ to $H$ in Table 2 and 68 saturated $OA(27, 3^{13}, 2)$ from Lam and Tonchev (1996), but also used an algorithm due to Xu (2002) to construct new designs directly. They presented many efficient and practically useful three-level nonregular designs with 18 and 27 runs for the dual objectives. Evangelaras, Koukouvinos and Lappas (2007) completely enumerated all nonisomorphic orthogonal arrays with 18 runs and 3 levels. Their results suggest that there are 4, 12, 10, 8, and 3 nonisomorphic $OA(18, 3^m, 2)$ for $m = 3$, 4, 5, 6, and 7, respectively. Evangelaras, Koukouvinos and Lappas (2008) further completely enumerated all nonisomorphic $OA(27, 3^m, 2)$ for $m = 3$–13 and identified 129 nonisomorphic saturated $OA(27, 3^{13}, 2)$.

Loeppky, Sitter and Tang (2007) proposed to rank two-level orthogonal designs based on the number of estimable models containing a subset of main effects and their associated two-factor interactions. They argued that by ranking designs in this way, the experimenter can directly assess the usefulness of the experimental plan for the purpose in mind. They presented catalogs of useful designs with 16, 20, 24, and 28 runs.

All these algorithmic constructions are limited to small run sizes ($\leq 32$) due to the existence of a large number of designs and the difficulty of determining whether two designs are isomorphic or equivalent. Katsaounisa and Dean (2008) gave a survey and evaluation of methods for determination of equivalence of factorial designs. Fang, Zhang and Li (2007) proposed an optimization algorithm for constructing generalized minimum aberration designs. It is not clear how effective their algorithm is for constructing large designs. Bulutoglu and Margot (2008) recently completely classified some orthogonal arrays of strength 3 up to 56 runs and of strength 4 up to 144 runs. However, these arrays have a small number of factors ($\leq 11$).

## 7.2 Optimality and Theoretical Results

A powerful tool in the study of regular designs is the complementary design technique. Every regular design can be determined by a unique complementary design. It is convenient to study the complementary design when it is small. Tang and Deng (1999) developed a complementary design theory for minimum $G_2$-aberration nonregular designs and Xu and Wu (2001) further developed a theory for generalized minimum aberration designs. The theory was extended by Ai and Zhang (2004b) for blocked nonregular designs and by Ai and He (2006) for nonregular designs with multiple groups of factors, including robust parameter designs. However, unlike in the regular case, a nonregular design can have none, one or more than one complementary designs; therefore, the complementary design theory for nonregular designs is less useful than that for regular designs.

Xu (2003) gave several sufficient conditions for a design to have minimum moment aberration and generalized minimum aberration among all possible designs. One sufficient condition is that for an orthogonal array of strength $t$ its projection onto any $t+1$ columns does not have repeated runs. For example, consider the $OA(18, 3^6, 2)$ given by columns $C$ to $H$ in Table 2. It is easy to verify that its projection onto any three columns does not have repeated runs. Thus, this design (and any of its projections) has minimum moment aberration and generalized minimum aberration among all possible designs. Another sufficient condition is that the numbers of coincidences between distinct rows are constant or differ by at most one. In other words, a design is optimal under the minimum moment aberration and generalized minimum aberration criteria if its design points are equally or nearly equally spaced over the design region. As an example, the $OA(12, 2^{11}, 2)$ given in Table 1 is optimal because the number of coincidences between any two distinct rows is 5. Generalizing this, Zhang, Fang, Li and Sudjianto (2005) proposed a majorization framework and showed that orthogonality, aberration and uniformity criteria can be unified by properly choosing combinatorial and exponential kernels.

Tang and Deng (2003) presented construction methods that yields maximum generalized resolution designs for 3, 4 and 5 factors and any run size $N$ that is a multiple of 4. Butler (2003, 2004) presented a number of construction results that allow minimum $G_2$-aberration designs to be found for many of the cases with $N =$16, 24, 32, 48, 64 and 96 runs. Butler (2005) further developed theoretical results and presented methods that allow generalized minimum aberration designs to be constructed for more than two levels. A key tool used by Butler (2003, 2004, 2005) is some identities that link the generalized wordlength patterns with moments of the inner products or Hamming distances between the rows. These identities can be derived easily from the generalized

Pless power moment identities developed by Xu (2003).

Xu (2005a) constructed several nonregular designs with 32, 64, 128, and 256 runs and 7–16 factors from the Nordstrom and Robinson code, a well-known nonlinear code in coding theory. These designs are better than regular designs of the same size in terms of resolution, aberration and projectivity. By using linear programming he showed that 13 nonregular designs have minimum $G_2$-aberration among all possible designs and seven orthogonal arrays have unique generalized wordlength patterns.

Tang (2006) studied the existence and construction of orthogonal arrays that are robust to nonnegligible two-factor interactions. Butler (2007) showed that foldover designs are the only (regular or nonregular) two-level factorial designs of resolution IV or more for $N$ runs and $N/3 \leq m \leq N/2$ factors. Yang and Butler (2007) studied two-level nonregular designs of resolution IV or more containing clear two-factor interactions and presented necessary and sufficient conditions for the existence of such designs. They gave many designs in concise grid representations for $N = 48$ up to 192 and $N$ being a multiple of 16.

Stufken and Tang (2007) completely classified all two-level orthogonal arrays with $t+2$ factors, strength $t$ and any run size. The key tool they used is the theory of $J$-characteristics developed by Tang (2001). Cheng, Mee and Yee (2008) studied the construction of second-order saturated orthogonal arrays of strength three $OA(N, 2^m, 3)$, which allows $N - m - 1$ two-factor interactions to be estimated besides $m$ main effects.

### 7.3 Nonregular designs constructed via quaternary codes

The construction of large regular designs is known to be very difficult (Xu 2007). The problem is even harder for nonregular designs. The construction via quaternary codes is relatively straightforward and can generate good large nonregular designs.

A quaternary code is a linear subspace over $Z_4 = \{0, 1, 2, 3\}$ (mod 4), the ring of integers modulus 4. A surprising breakthrough in coding theory is that many famous nonlinear codes such as the Nordstrom and Robinson code can be constructed via quaternary codes (Hammons et al. 1994). A key device is the so-called Gray map:

$$\phi : 0 \to (0,0), 1 \to (0,1), 2 \to (1,1), 3 \to (1,0),$$

which maps each symbol in $Z_4$ to a pair of symbols in $Z_2$. Let $G$ be a $k \times n$ matrix and let $C$ consist of all possible linear combinations of the row vectors of $G$ over $Z_4$. Applying the Gray map to $C$, one obtains a $4^k \times 2n$ matrix or a two-level design, denoted by $D$. Although $C$ is linear over $Z_4$, $D$ may or may not be linear over $Z_2$.

Xu and Wong (2007) described a systematic procedure for constructing nonregular designs from quaternary codes. They first generated a $k \times (4^k - 2^k)/2$ generator matrix $G$ which has the following properties: (i) it does not have any column containing entries 0 and 2 only and (ii) none of its column is a multiple of another column over $Z_4$. Xu and Wong (2007) showed that the binary image $D$ generated by $G$ is a $4^k \times (4^k - 2^k)$ design with resolution 3.5 whereas regular designs of the same size have resolution 3. To obtain designs with less than $4^k - 2^k$ columns, they developed a sequential algorithm, similar to those by Chen, Sun and Wu (1993) and Xu (2005b). They also presented a collection of nonregular designs with 32, 64, 128 and 256 runs and up to 64 factors, many of which are better than regular designs of the same size in terms of resolution, aberration and projectivity.

Phoa and Xu (2008) further investigated the properties of quarter-fraction designs which can be defined by a generator matrix that consists of an identity matrix plus an extra column. They showed that the resolution, wordlength and projectivity can be calculated in terms of the frequencies of the numbers 1, 2 and 3 that appear in the extra column. These results enabled them to construct optimal quarter-fraction designs via quaternary codes under the maximum resolution, minimum aberration and maximum projectivity criteria. These designs are often better than regular designs of the same size in terms of the design criterion. The generalized minimum aberration designs constructed via quaternary codes have the same aberration as the minimum aberration regular designs, and frequently with larger resolution and projectivity. A maximum projectivity design is often different from a minimum aberration or maximum resolution design but can have much larger projectivity than a minimum aberration regular design. They further showed that some of these designs have generalized minimum aberration and maximum projectivity among all possible designs.

There are two obvious advantages of using quaternary codes to construct nonregular designs: (1) relatively straightforward construction and (2) simple design representation. Since the designs are constructed via linear codes over $Z_4$, one can use column indexes to describe these designs. The linear structure of a quaternary code also facilitates the derivation and analytical study of properties of nonregular designs.

# 8   Concluding Remarks and Future Directions

We have discussed recent developments in nonregular fractional factorial designs in the preceding sections. In a nutshell, when we compare regular designs with nonregular designs, nonregular designs have the following advantages:

1. require smaller run size

2. are more flexible in accommodating a variate number of factor levels

3. have better geometrical or hidden projection properties

4. have higher generalized resolution and projectivity

5. have less generalized aberration

6. lessen the contamination of nonnegligible two-factor interactions on the estimation of the main effects

Some of the disadvantages of nonregular designs are that they are more complicated to analyze and some estimates of factorial effects may have larger variance than others.

This review does not include the developments in supersaturated designs, which are factorial designs whose run sizes are not enough for estimating all the main effects. The research on supersaturated designs has been very active since the influential work of Lin (1993) and Wu (1993). Broadly speaking, supersaturated designs are nonregular designs and optimality criteria such as generalized resolution and generalized minimum aberration can also be applied directly. As mentioned earlier, Xu (2003) developed a unified theory for nonregular and supersaturated designs using the concept of minimum moment aberration. Xu and Wu (2005) obtained more theoretical results under the generalized minimum aberration criterion for multi-level and mixed-level supersaturated designs. Gilmour (2006) reviewed the recent development of two-level supersaturated designs for factor screening.

Finally we highlight some future directions of research for nonregular designs and comment briefly why we feel they are useful:

1. applications of nonregular designs

2. analysis of nonregular designs

3. construction of good nonregular designs with large run sizes

4. optimality results with respect to the generalized resolution.

Despite significant developments in recent years and the advantages of using nonregular designs, they are still widely used for screening main effects only in practice and applications are largely limited to industry. We hope that by documenting recent advances in nonregular designs, our work may stimulate greater research interest in nonregular designs. We feel that there are opportunities that nonregular designs can be effectively applied to other fields to reduce experimental cost and gain improvement in statistical efficiency.

The analysis of nonregular designs requires more attention. Although one can use any general purpose variable selection procedures, it is desirable to have user-friendly packages that incorporate the special features of nonregular designs in the analysis. More analysis strategies and comparisons are needed to further understand and utilize the complex aliasing structure of nonregular designs. Different procedures do not always lead to unequivocal conclusions. When that happens, extra runs are needed to resolve the ambiguity. See Meyer, Steinberg and Box (1996) and Box, Hunter and Hunter (2005, Chapter 7) for methods of constructing follow-up designs.

There are plenty of catalogs of optimal nonregular designs with small run sizes ($\leq 32$). With the popularity of computer experiments, more and more large nonregular factorial designs will be used in practice. Mee (2004) illustrated how nonregular designs can be used to reduce run sizes significantly in applications requiring estimation of the main effects and two-factor interactions for a large number of factors. He suggested the use of a 2,096-run nonregular design with 47 factors as an alternative to a 4,096-run regular designs in a ballistic missile simulation application. The quaternary code construction method is very promising in this regard and is able to produce large nonregular designs with good properties. Indeed, this 2,096-run nonregular design can be conveniently constructed via quaternary codes.

Several optimality results and theories have been obtained for the minimum $G_2$-aberration and the generalized minimum aberration. However, at the present time, there is very limited results on the generalized resolution for nonoregular designs. This is useful because it is always helpful to know whether a design is close to the optimal design or not, without knowing the optimal design.

# Acknowledgments

All authors were supported in part by National Institute of Health grant 5R01 GM072876. The research of Xu and Phoa was also partially supported by a National Science Foundation grant

DMS-0806137. The authors are grateful to C. F. J. Wu and Boxin Tang for their comments and helps on the history of the developments in nonregular designs.# References

1. Ai, M. Y. and He, S.Y. (2006). Generalized wordtype pattern for nonregular factorial designs with multiple groups of factors. *Metrika*, **64**, 95-108.

2. Ai, M. Y. and Zhang, R. C. (2004a). Projection justification of generalized minimum aberration for asymmetrical fractional factorial designs. *Metrika*, **60**, 279-285.

3. Ai, M. Y. and Zhang, R. C. (2004b). Theory of optimal blocking of nonregular factorial designs. *Canad. J. Statist.*, **32**, 57-72.

4. Ai, M. Y., Li, P. F. and Zhang, R. C. (2005). Optimal criteria and equivalence for nonregular fractional factorial designs. *Metrika*, **62**, 73-83.

5. Box, G. E. P. and Hunter, J. S. (1961). The $2^{k-p}$ fractional factorial designs. *Technometrics*, **3**, 311–351, 449–458.

6. Box, G. E. P. and Meyer, R. D. (1993). Finding the active factors in fractionated screening experiments. *J. Quality Technology*, **25**, 94–105.

7. Box, G. E. P. and Tyssedal, J. (1996). Projective properties of certain orthogonal arrays. *Biometrika*, **83**, 950–955.

8. Box, G. E. P., Hunter, W. G. and Hunter, J. S. (2005). *Statistics for Experimenters: Design, Innovation, and Discovery*, 2nd ed. New York: Wiley.

9. Bulutoglu, D. A. and Cheng, C. S. (2003). Hidden projection properties of some nonregular fractional factorial designs and their applications. *Ann. Statist.*, **31**, 1012-1026.

10. Bulutoglu, D. A. and Margot, F. (2008). Classification of orthogonal arrays by integer programming. *J. Statist. Plann. Inference*, **138**, 654–666.

11. Butler, N. A. (2003). Minimum aberration construction results for nonregular two-level fractional factorial designs. *Biometrika*, **90**, 891–898.

12. Butler, N. A. (2004). Minimum $G_2$-aberration properties of two-level foldover designs. *Statist. Probab. Lett.*, **67**, 121–132.

13. Butler, N. A. (2005). Generalised minimum aberration construction results for symmetrical orthogonal arrays. *Biometrika*, **92**, 485-491.

14. Butler, N. A. (2007). Results for two-level fractional factorial designs of resolution IV or more. *J. Statist. Plann. Inference*, **137**, 317-323.
24

15. Chen, J., Sun, D. X. and Wu, C. F. J. (1993). A catalogue of two-level and three-level fractional factorial designs with small runs. *Internat. Statist. Rev.*, **61**, 131–145.

16. Cheng, C. S. (1980). Orthogonal arrays with variable numbers of symbols. *Ann. Statist.*, **8**, 447–453.

17. Cheng, C. S. (1995). Some projection properties of orthogonal arrays. *Ann. Statist.*, **23**, 1223–1233.

18. Cheng, C. S. (1998). Some hidden projection properties of orthogonal arrays with strength three. *Biometrika*, **85**, 491–495.

19. Cheng, C. S. (2006). Projection properties of factorial designs for factor screening. *Screening: Methods for Experimentation in Industry, Drug Discovery, and Genetics*, Ed. A. Dean and S. Lewis, pp. 156–168. New York: Springer.

20. Cheng, C. S., Deng, L. Y. and Tang, B. (2002). Generalized minimum aberration and design efficiency for nonregular fractional factorial designs. *Statist. Sinica*, **12**, 991–1000.

21. Cheng, C. S., Mee, R. W. and Yee, O. (2008). Second order saturated orthogonal arrays of strength three. *Statist. Sinica*, **18**, 105-119.

22. Cheng, C. S., Steinberg, D. M. and Sun, D. X. (1999). Minimum aberration and model robustness for two-level fractional factorial designs. *J. Roy. Statist. Soc. Ser. B*, **61**, 85–93.

23. Cheng, S. W. and Wu, C. F. J. (2001). Factor screening and response surface exploration (with discussion). *Statist. Sinica*, **11**, 553–604.

24. Cheng, S. W. and Ye, K. Q. (2004). Geometric isomorphism and minimum aberration for factorial designs with quantitative factors. *Ann. Statist.*, **32**, 2168–2185.

25. Cheng, S. W., Li, W. and Ye, K. Q. (2004). Blocked nonregular two-level factorial designs. *Technometrics*, **46**, 269–279.

26. Chipman, H., Hamada, M. and Wu, C.F.J. (1997). A Bayesian variable-selection approach for analyzing designed experiments with complex aliasing. *Technometrics*, **39**, 372–381.

27. Dean, A. M. and Voss, D. T. (1999). *Design and analysis of experiments*. New York: Springer.

28. Deng, L. Y. and Tang, B. (1999). Generalized resolution and minimum aberration criteria for Plackett-Burman and other nonregular factorial designs. *Statist. Sinica*, **9**, 1071–1082.

29. Deng, L. Y. and Tang, B. (2002). Design selection and classification for Hadamard matrices using generalized minimum aberration criteria. *Technometrics*, **44**, 173–184.

30. Dey, A. (2005). Projection properties of some orthogonal arrays. *Statist. Probab. Lett.*, **75**, 298–306.

31. Efron, B., Hastie, T., Johnstone, I. and Tibshirani, R. (2004). Least angle regression. *Ann. Statist.*, **32**, 407-499.

32. Evangelaras, H., Koukouvinos, C. and Lappas, E. (2007). 18-run nonisomorphic three level orthogonal arrays. *Metrika*, **66**, 31-37.

33. Evangelaras, H., Koukouvinos, C. and Lappas, E. (2008). 27-run nonisomorphic three level orthogonal arrays: Identification, evaluation and projection properties. *Utilitas Mathematica*, to appear.

34. Evangelaras, H., Koukouvinos, C., Dean, A. M., and Dingus, C. A. (2005). Projection properties of certain three level orthogonal arrays. *Metrika*, **62**, 241–257.

35. Fang, K. T. and Mukerjee, R. (2000). A connection between uniformity and aberration in regular fractions of two-level factorials. *Biometrika*, **87**, 193–198.

36. Fang, K. T. and Qin, H. (2005). Uniformity pattern and related criteria for two-level factorials. *Science in China, Series A: Mathematics*, **48**, 1-11.

37. Fang, K. T. and Wang, Y. (1994). *Number-theoretic methods in statistics*. London: Chapman and Hall.

38. Fang, K. T., Li, R. and Sudjianto, A. (2006). *Design and Modeling for Computer Experiments*. London: Chapman and Hall/CRC.

39. Fang, K. T., Lin, D. K. J., Winker, P. and Zhang, Y. (2000). Uniform design: Theory and application. *Technometrics*, **42**, 237-248.

40. Fang, K. T., Zhang, A. and Li, R. (2007). An effective algorithm for generation of factorial designs with generalized minimum aberration. *J. Complexity*, **23**, 740–751.

41. Fries, A. and Hunter, W. G. (1980). Minimum aberration $2^{k-p}$ designs. *Technometrics*, **22**, 601–608.

42. Gilmour, S. (2006). Factor screening via supersaturated designs. *Screening: Methods for Experimentation in Industry, Drug Discovery, and Genetics*, Ed. A. Dean and S. Lewis, pp. 169–190. New York: Springer.

43. Hall, M. Jr. (1961). Hadamard matrix of order 16. *Jet Propulsion Laboratory Research Summary*, **1**, 21–26.

44. Hall, M. Jr. (1965). Hadamard matrix of order 20. *Jet Propulsion Laboratory Technical Report*, **1**, 32–76.

45. Hamada, M. and Wu, C. F. J. (1992). Analysis of designed experiments with complex aliasing. *J. Quality Technology*, **24**, 130–137.

46. Hammons, A. R., Jr., Kumar, P. V., Calderbank, A. R., Sloane, N. J. A. and Sole, P. (1994). The $Z_4$-linearity of Kerdock, Preparata, Goethals, and related codes. *IEEE Trans. Inform. Theory*, **40**, 301–319.

47. Hedayat, A. S., Sloane, N. J. A. and Stufken, J. (1999). *Orthogonal Arrays: Theory and Applications*. New York: Springer.

48. Hickernell, F. J. (1998) A generalized discrepancy and quadrature error bound. *Math. Comp.*, **67**, 299–322.

49. Hickernell, F. J. and Liu, M. Q. (2002), Uniform designs limit aliasing. *Biometrika*, **89**, 893–904.

50. Hunter, G. B., Hodi, F. S. and Eager, T. W. (1982). High-cycle fatigue of weld repaired cast Ti-6Al-4V. *Metallurgical Transactions A*, **13**, 1589–1594.

51. Katsaounis, T. I. and Dean, A. M. (2008). A survey and evaluation of methods for determination of combinatorial equivalence of factorial designs. *J. Statist. Plann. Inference*, **138**, 245-258.

52. King, C. and Allen, L. (1987). Optimization of winding operation for radio frequency chokes. *Fifth Symposium on Taguchi Methods,* pp. 67–80. Dearborn, Michigan: American Supplier Institute.

53. Kuhfeld, W. F. (2005). *Marketing Research Methods in SAS*. SAS Institute Inc., Cary, NC. Available at http://support.sas.com/techsup/technote/ts722.pdf.

54. Lam, C. and Tonchev, V. D. (1996). Classification of affine resolvable 2-(27,9,4) designs. *J. Statist. Plann. Inference*, **56**, 187–202.

55. Li, W. (2006). Screening designs for model selection. *Screening: Methods for Experimentation in Industry, Drug Discovery, and Genetics*, Ed. A. Dean and S. Lewis, pp. 207-234. New York: Springer.

56. Li, W., Lin, D. K. J. and Ye, K. Q. (2003). Optimal foldover plans for two-level nonregular orthogonal designs. *Technometrics*, **45**, 347-351.

57. Li, Y., Deng, L.-Y. and Tang, B. (2004). Design catalog based on minimum $G$-aberration. *J. Statist. Plann. Inference*, **124**, 219–230.

58. Lin, D. K. J. (1993). A new class of supersaturated designs. *Technometrics*, **35**, 28–31.

59. Lin, D. K. J. and Draper, N. R. (1992). Projection properties of Plackett and Burman designs. *Technometrics*, **34**, 423–428.

60. Liu, M. Q., Fang, K. T. and Hickernell, F. J. (2006). Connections among different criteria for asymmetrical fractional factorial designs. *Statist. Sinica*, **16**, 1285-1297.

61. Loeppky, J. L., Bingham, D. and Sitter R.R. (2006). Constructing non-regular robust parameter designs. *J. Statist. Plann. Inference*, **136**, 3710-3729.

62. Loeppky, J. L., Sitter, R. R. and Tang, B. (2007). Nonregular designs with desirable projection properties. *Technometrics*, **49**, 454-467.

63. Ma, C. X. and Fang, K. T. (2001). A note on generalized aberration in factorial designs. *Metrika*, **53**, 85–93.

64. MacWilliams, F. J. and Sloane, N. J. A. (1977). *The Theory of Error-correcting Codes*. Amsterdam: North-Holland.

65. Mandal, A. and Mukerjee, R. (2005). Design efficiency under model uncertainty for nonregular fractions of general factorials. *Statist. Sinica*, **15**, 697-707.

66. Mee, R. W. (2004). Efficient two-level designs for estimating main effects and two-factor interactions. *J. Quality Technology*, **36**, 400–412.

67. Meyer, R. D., Steinberg, D. M. and Box, G. (1996). Follow-up designs to resolve confounding in multifactor experiments. *Technometrics*, **38**, 303-313.

68. Montgomery, D. C. (2005). *Design and analysis of experiments*. 6th ed. New York: Wiley.

69. Mukerjee, R. and Wu, C. F. J. (2006). *A Modern Theory of Factorial Designs*. New York: Springer.

70. Paley, R. E. A. C. (1933). On orthogonal matrices. *J. Math. Phys.*, **12**, 311–320.

71. Phoa, F. K. H. and Xu, H. (2008). Quarter-fraction factorial designs constructed via quaternary codes. UCLA Statistics Electronic Publications, preprint 538. `http://preprints.stat.ucla.edu/`. Accepted by *Ann. Statist.*

72. Phoa, F. K. H., Pan, Y.-H. and Xu, H. (2007). Analysis of supersaturated designs via the Dantzig selector. UCLA Statistics Electronic Publications, preprint 526. `http://preprints.stat.ucla.edu/`. Accepted by *J. Statist. Plann. Inference.*

73. Plackett, R. L. and Burman, J. P. (1946). The design of optimum multifactorial experiments. *Biometrika*, **33**, 305–325.

74. Qin, H. and Ai, M. (2007). A note on the connection between uniformity and generalized minimum aberration. *Statistical Papers*, **48**, 491-502.

75. Qin, H. and Fang, K. T. (2004). Discrete discrepancy in factorial designs. *Metrika*, **60**, 59-72.

76. Qin, H., Zou, N. and Chatterjee, K. (2008). Connection between uniformity and minimum moment aberration. *Metrika*, available online, DOI 10.1007/s00184-008-0180-9.

77. Rao, C. R. (1947). Factorial experiments derivable from combinatorial arrangements of arrays. *J. Roy. Statist. Soc. Ser. B*, **9**, 128–139.

78. Rao, C. R. (1973). Some combinatorial problems of arrays and applications to design of experiments. *Survey of combinatorial theory*, Ed. J. N. Srivastava, pp. 349–359. Amsterdam: North-Holland.

79. Stufken, J. and Tang, B. (2007). Complete enumeration of two-level orthogonal arrays of strength $d$ with $d+2$ constraints. *Ann. Statist.*, **35**, 793–814.

80. Sun, D. X. and Wu, C. F. J. (1993). Statistical properties of Hadamard matrices of order 16. *Quality Through Engineering Design*, Ed. W. Kuo, pp. 169–179. New York: Elsevier.

81. Sun, D. X., Li, W. and Ye, K. Q. (2002). An algorithm for sequentially constructing nonisomorphic orthogonal designs and its applications. Technical report SUNYSB-AMS-02-13, Department of Applied Mathematics and Statistics, SUNY at Stony Brook.

82. Taguchi, G. (1987). *System of Experimental Design*. White Plain, New York: UNIPUB.

83. Tang, B. (2001). Theory of *J*-characteristics for fractional factorial designs and projection justification of minimum $G_2$-aberration. *Biometrika*, **88**, 401–407.

84. Tang, B. (2006). Orthogonal arrays robust to nonnegligible two-factor interactions. *Biometrika*, **93**, 137–146.

85. Tang, B. and Deng, L. Y. (1999). Minimum $G_2$-aberration for non-regular fractional factorial designs. *Ann. Statist.*, **27**, 1914–1926.

86. Tang, B. and Deng, L.-Y. (2003). Construction of generalized minimum aberration designs of 3, 4 and 5 factors. *J. Statist. Plann. Inference*, **113**, 335–340.

87. Tsai, P.-W., Gilmour, S. G. and Mead, R. (2000). Projective three-level main effects designs robust to model uncertainty. *Biometrika*, **87**, 467–475.

88. Tsai, P.-W., Gilmour, S. G. and Mead, R. (2004). Some new three-level orthogonal main effects plans robust to model uncertainty. *Statist. Sinica*, **14**, 1075–1084.

89. Wang, J. C. and Wu, C. F. J. (1995). A hidden projection property of Plackett-Burman and related designs. *Statist. Sinica*, **5**, 235–250.

90. Westfall, P.H., Young, S.S. and Lin, D.K.J. (1998). Forward selection error control in the analysis of supersaturated designs. *Statist. Sinica*, **8**, 101-117.

91. Wu, C. F. J. (1993). Construction of supersaturated designs through partially aliased interactions. *Biometrika*, **80**, 661–669.

92. Wu, C. F. J. and Hamada, M. (2000). *Experiments: Planning, Analysis and Parameter Design Optimization*. New York: Wiley.

93. Xu, H. (2002). An algorithm for constructing orthogonal and nearly-orthogonal arrays with mixed levels and small runs. *Technometrics*, **44**, 356–368.

94. Xu, H. (2003). Minimum moment aberration for nonregular designs and supersaturated designs. *Statist. Sinica*, **13**, 691–708.

95. Xu, H. (2005a). Some nonregular designs from the Nordstrom and Robinson code and their statistical properties. *Biometrika*, **92**, 385–397.

96. Xu, H. (2005b). A catalogue of three-level regular fractional factorial designs. *Metrika*, **62**, 259–281.

97. Xu, H. (2006). Blocked regular fractional factorial designs with minimum aberration. *Ann. Statist.*, **34**, 2534–2553.

98. Xu, H. (2007). Algorithmic construction of efficient fractional factorial designs with large run sizes. UCLA Statistics Electronic Publications, preprint 520. `http://preprints.stat.ucla.edu/`.

99. Xu, H. and Deng, L. Y. (2005). Moment aberration projection for nonregular fractional factorial designs. *Technometrics*, **47**, 121–131.

100. Xu, H. and Lau, S. (2006). Minimum aberration blocking schemes for two- and three-level fractional factorial designs. *J. Statist. Plann. Inference*, **136**, 4088–4118.

101. Xu, H. and Wong, A. (2007). Two-level nonregular designs from quaternary linear codes. *Statist. Sinica*, **17**, 1191–1213.

102. Xu, H. and Wu, C. F. J. (2001). Generalized minimum aberration for asymmetrical fractional factorial designs. *Ann. Statist.*, **29**, 1066–1077.

103. Xu, H. and Wu, C. F. J. (2005). Construction of optimal multi-level supersaturated designs. *Ann. Statist.*, **33**, 2811–2836.

104. Xu, H., Cheng, S.W. and Wu, C. F. J. (2004). Optimal projective three-level designs for factor screening and interaction detection. *Technometrics*, **46**, 280–292.

105. Yang, G. and Butler, N. A. (2007). Nonregular two-level designs of resolution IV or more containing clear two-factor interactions. *Statist. Probab. Lett.*, **77**, 566-575.

106. Ye, K. Q. (2003). Indicator functions and its application in two-level factorial designs. *Ann. Statist.*, **31**, 984–994.

107. Ye, K. Q. (2004). A note on regular fractional factorial designs. *Statist. Sinica*, **14**, 1069–1074.

108. Yuan, M., Joseph, V. R. and Lin, Y. (2007). An efficient variable selection approach for analyzing designed experiments. *Technometrics*, **49**, 430-439.

109. Zhang, A., Fang, K. T., Li, R. and Sudjianto, A. (2005). Majorization framework for balanced lattice designs. *Ann. Statist.*, **33**, 2837-2853.

Department of Statistics, University of California, Los Angeles, CA 90095, U.S.A.

E-mail: hqxu@stat.ucla.edu

Department of Statistics, University of California, Los Angeles, CA 90095, U.S.A.

E-mail: fredphoa@stat.ucla.edu

Department of Biostatistics, University of California, Los Angeles, CA 90095, U.S.A.

E-mail: wkwong@ucla.edu